\documentclass[12pt]{article}

\usepackage[left=1in,top=1in,right=1in,bottom=1in,head=.1in,nofoot]{geometry}
\usepackage{setspace,url,bm,amsmath} 
\usepackage{titlesec} 
\titlelabel{\thetitle.\quad} 
\usepackage[margin=20pt]{subcaption}
\usepackage{amsmath,amsfonts,amsthm,amssymb}
\usepackage{graphicx}
\usepackage[colorlinks=true,linkcolor=black,citecolor=blue,urlcolor=blue]{hyperref}
\usepackage[table]{xcolor}
\usepackage{multirow}
\usepackage{array}
\usepackage{booktabs}
\usepackage{threeparttable}
\usepackage{comment}
\usepackage{float}
\usepackage{natbib}
\usepackage{indentfirst}

\newtheorem*{theorem*}{Theorem}

\newtheorem{example}{Example}

\newtheorem*{corollary*}{Corollary}

\def\nt{n_1}
\def\nc{n_0}

\usepackage[ruled,vlined]{algorithm2e}
\usepackage{booktabs,multirow,caption,threeparttable}

\usepackage{enumitem}
\setenumerate[1]{itemsep=0pt,partopsep=0pt,parsep=\parskip,topsep=5pt}
\setitemize[1]{itemsep=0pt,partopsep=0pt,parsep=\parskip,topsep=5pt}

\newenvironment{examplecont}
{\addtocounter{example}{-1}
 
\begin{example}}
{\end{example}}

\def\cZ{\mathcal{Z}}
\def\z{\boldsymbol{z}}
\def\zp{\boldsymbol{z}_{\pi}}
\def\zpp{\boldsymbol{z}_{\pi'}}
\def\d{\boldsymbol{\delta}}
\def\Y{\boldsymbol{Y}}
\def\tY{\widetilde{\boldsymbol{Y}}}

\def\nt{n_1}
\def\nc{n_0}

\def\t{t}
\def\s{s^2}

\begin{document}
\begin{singlespace}

\title{\bf Rejoinder to Reader Reaction ``On exact randomization-based covariate-adjusted confidence intervals" by Jacob Fiksel}
\author{
Ke Zhu\\
Department of Statistics, North Carolina State University\\
Department of Biostatistics and Bioinformatics, Duke University\\
Hanzhong Liu\thanks{Corresponding author: lhz2016@tsinghua.edu.cn}\\
Center for Statistical Science, Department of Industrial Engineering,\\ Tsinghua University
}

\date{}
\maketitle
\end{singlespace}



\setcounter{page}{1}



\section{Introduction}\label{sec:intro}

We applaud \citet{fiksel2024} for their valuable contributions to randomization-based inference, particularly their work on inverting the Fisher randomization test (FRT) to construct confidence intervals using the covariate-adjusted test statistic. 
FRT is advocated by many scholars because it produces finite-sample exact $p$-values for any test statistic and can be easily adopted for any experimental design \citep{rosenberger2019randomization, proschan2019re,  young2019, bind2020}. 
By inverting FRTs, we can construct the randomization-based confidence interval (RBCI). However, there are two main criticisms for the randomization-based inference with FRT and RBCI.

First, FRT is computationally intensive, although this issue is gradually diminishing with the advancement of modern computing capabilities \citep{bind2020}. More importantly, RBCIs established through existing approximation approaches generally lack theoretical guarantees for achieving the desired coverage probability \citep{luo2021}.
\citet{luo2021} conducted a formal investigation on the theoretical properties of the RBCI generated by the bisection method and provided the theoretical guarantee of the desired coverage probability under a monotonic condition for the $p$-value function produced by the test statistic.
When the test statistic does not satisfy this monotonic condition, the RBCI generated by approximation methods is questionable.

Another criticism of the FRT is that it is valid for testing the sharp null hypothesis rather than the weak null hypothesis. Correspondingly, the exact coverage of the RBCI also requires the restricted condition of constant treatment effects. \citet{wu2020} proved that FRT using studentized statistic is also asymptotically valid for testing weak null hypothesis and the corresponding RBCI asymptotically achieves nominal coverage probability without the condition of constant treatment effects. However, the studentized statistic, as many other commonly used test statistics, does not satisfy the monotonic condition in \citet{luo2021}. This critical contradiction has also hindered the application of RBCI in practice. Moreover, the coverage probability of RBCI may exceed the nominal level due to the discreteness of the $p$-value function, especially when the sample size is small.

To the best of our knowledge, \citet{zhu2023pair} is the first paper analytically inverting the FRT for the difference-in-means statistic. \citet{fiksel2024} extended this analytical approach to the covariate-adjusted statistic, which produced a monotonic $p$-value function under certain conditions. 
In this rejoinder, we propose an analytical approach to invert the FRT for test statistics that produce a non-monotonic $p$-value function, with the studentized $t$-statistic as an important special case. Exploiting our analytical approach, we can recover the non-monotonic $p$-value function and construct RBCI based on the studentized $t$-statistic. The RBCI generated by the proposed analytical approach is guaranteed to achieve the desired coverage probability and resolve the contradiction between \citet{luo2021} and \citet{wu2020}. 
Simulation results validate our findings and demonstrate that our method is also computationally efficient.

\section{Exact randomization-based confidence interval based on studentized \textit{t}-statistic }
\label{sec:method}

Let $Y_i(0)$ and $Y_i(1)$ denote the potential outcomes of unit $i$, $i=1,\ldots,n$.
Let $\cZ$ denote all possible assignments for the experiment.
Let $\z=(z_1,\ldots,z_n)\in \cZ$ ($z_i \in \{0,1\}$) denote the observed assignment vector. 
Let $\Y=(Y_1,\ldots,Y_n)$ denote the observed outcome with $Y_i=z_iY_i(1)+(1-z_i)Y_i(0)$.

Under $H_0^\theta:Y_i(1)-Y_i(0)=\theta$, the imputed potential outcomes are
$$
\begin{aligned}
& \widetilde{Y}_i(1)= \begin{cases}Y_i & \text { if } z_i=1 \\
Y_i+\theta & \text { if } z_i=0\end{cases} ,\\
& \widetilde{Y}_i(0)= \begin{cases}Y_i & \text { if } z_i=0 \\
Y_i-\theta & \text { if } z_i=1 \end{cases}.
\end{aligned}
$$
Let $\zp=(z_{\pi,1},\ldots,z_{\pi,n})\in\cZ$ denote an alternative assignment vector. 
The outcome under $\zp$ is 
$$
\widetilde{Y}_i= 
z_{\pi,i}\widetilde{Y}_i(1)+(1-z_{\pi,i})\widetilde{Y}_i(0)
=\begin{cases}Y_i & \text { if } z_{\pi,i}=z_{i} \\
Y_i+\theta & \text { if } z_{\pi,i}=1,z_{i}=0\\
Y_i-\theta & \text { if } z_{\pi,i}=0,z_{i}=1
\end{cases}.
$$
A simplified expression for $\tY=(\widetilde{Y}_1,\ldots,\widetilde{Y}_n)$ is
$$
\tY=\boldsymbol{Y}+\d\theta,
$$
where $\d=(\delta_1,\ldots,\delta_n)=\zp-\z$ denotes the difference between $\zp$ and $\z$.

\subsection{Step 1: Determining jump points}
\label{sec:1}

Consider the alternative hypothesis $H_1^{\theta+}:Y_i(1)-Y_i(0)>\theta$ and any test statistic $T(\z,\Y)$. The $p$-value function is defined as
$$
p^+(\theta)=\frac{1}{|\cZ|} \sum_{\zp\in\cZ}\mathbf{1}\{T(\zp,\tY)\ge T(\z,\Y)\},
$$
where $|\cZ|$ is the number of assignments in $\cZ$ and $\mathbf{1}\{\cdot\}$ is the indicator function. Similarly, we can define $p^-(\theta)$ for $H_1^{\theta-}:Y_i(1)-Y_i(0)<\theta$.
The $p$-value function may not be monotonic, but it is always a step function with jump points that could be solved by a basic equation:
$$
T(\z,\Y)=T(\zp,\tY)\equiv T(\zp,\boldsymbol{Y}+\d\theta).
$$

Consider the difference-in-means statistic:
$$
\widehat{\tau}(\z,\Y)=\frac{1}{\nt} \sum_{i=1}^n z_i Y_i  -\frac{1}{\nc} \sum_{i=1}^n \left(1-z_i\right)Y_i,
$$
where $\nt=\sum_{i=1}^n z_i$ and $\nc=\sum_{i=1}^n (1-z_i)$ denote the number of units in the treatment and control groups, respectively.
By replacing $(\z,\Y)$ with $(\zp,\tY)$, we can similarly define the value of the statistic under $\zp$ by $\widehat{\tau}(\zp,\tY)$. Since $\tY=\boldsymbol{Y}+\d\theta$, by some algebra, we have
$$
\widehat{\tau}(\zp,\tY)=\widehat{\tau}(\zp,\Y)+\widehat{\tau}(\zp,\d)\theta.
$$
Then, we solve the basic equation $\widehat{\tau}(\z,\Y)=\widehat{\tau}(\zp,\tY)$ and obtain the root
$$
\theta_{\pi}=\frac{\widehat{\tau}(\z,\Y)-\widehat{\tau}(\zp,\Y)}{\widehat{\tau}(\zp,\d)}.
$$
This solution is equivalent to that obtained in \citet{zhu2023pair}. However, the above expressions are more concise and motivate the following solution for the studentized $t$-statistic.

Consider the studentized $t$-statistic :
$$
\t(\z,\Y)=\frac{\widehat{\tau}(\z,\Y)}{\sqrt{\s(\z,\Y,\Y)}},
$$
where
\begin{align*}
   \s(\z,\Y,\Y)=&\frac{1}{\nt}\frac{1}{\nt-1}\sum_{i=1}^n z_i \left(Y_i-\frac{1}{\nt} \sum_{i=1}^n z_i Y_i \right)^2 +\frac{1}{\nc}\frac{1}{\nc-1}\sum_{i=1}^n (1-z_i) \left(Y_i-\frac{1}{\nc} \sum_{i=1}^n (1-z_i) Y_i \right)^2.
\end{align*}
By replacing $(\z,\Y)$ with $(\zp,\tY)$, we can similarly define $\s(\zp,\tY,\tY)$ and $\t(\zp,\tY)$. Since $\tY=\boldsymbol{Y}+\d\theta$, by some algebra, we have
$$
\t(\zp,\tY)=\frac{\widehat{\tau}(\zp,\tY)}{\sqrt{\s(\zp,\tY,\tY)}}=
\frac{
\widehat{\tau}(\zp,\Y)+\widehat{\tau}(\zp,\d)\theta
}{\sqrt{
\s(\zp,\Y,\Y)+2\s(\zp,\Y,\d)\theta + \s(\zp,\d,\d) \theta^2
}}.
$$
Then, we can transform the basic equation $\t(\z,\Y)=\t(\zp,\tY)$ into a quadratic equation
\begin{align*}
&\left\{
\t^2(\z,\Y)\s(\zp,\d,\d)-\widehat{\tau}^2(\zp,\d)
\right\}\theta_{\pi}^2
\\+2&\left\{
\t^2(\z,\Y)\s(\zp,\Y,\d)-\widehat{\tau}(\zp,\Y)\widehat{\tau}(\zp,\d)
\right\}\theta_{\pi}
\\
+&\left\{
\t^2(\z,\Y)\s(\zp,\Y,\Y)-\widehat{\tau}^2(\zp,\Y)
\right\}=0,
\end{align*}
whose roots have explicit forms.

We solve $\theta_{\pi}$ for every $\zp\in\cZ$ and denote the set by $\{\theta_{\pi}: \zp\in\cZ\}$, which contains all jump points of the $p$-value function. Once we have determined all the jump points, one way to recover the $p$-value function is by evaluating it at these jump points, which is feasible but time-consuming. Fortunately, we can recover the $p$-value function more efficiently, even when it is non-monotonic.

\begin{figure}[p]
\centering
    \includegraphics[width = 0.75\textwidth]{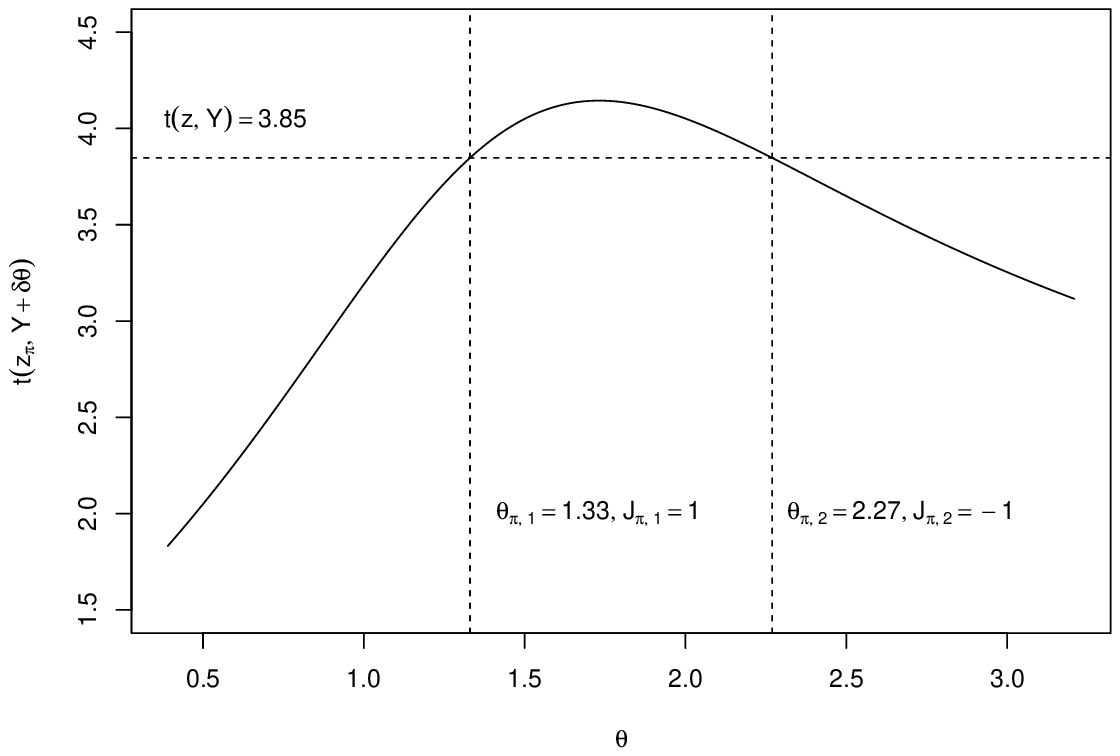}
\caption{The value of the studentized $t$-statistic $\t(\zp,\Y+\d\theta)$ under $\zp$. The horizontal dashed line is the observed value of the studentized $t$-statistic $\t(\z,\Y)$. The vertical dashed lines are the roots of the basic equation $\t(\z,\Y)=\t(\zp,\Y+\d\theta)$. This Figure corresponds to Example \ref{ex:1} with 8 units.}
\label{fig:t}
\end{figure}

\begin{example}
\label{ex:1}
    We revisit the example in \citet{luo2021}. The potential outcomes are $\Y(0)=(0.14, 1.12, 0.80, 1.80, 0.90, 0.44, 1.13, 0.53)$ and $\Y(1)=\Y(0)+1$. We randomly assign $4$ units to the treatment group and $4$ units to the control group. The observed assignment is $z=(1, 1, 0, 1, 0, 0, 1, 0)$. Consider an alternative assignment vector $\zp=(1,1,1,1,0,0,0,0)$.
    Consider the studentized $t$-statistic as the test statistic. We plot the value of $\t(\zp,\tY)=\t(\zp,\Y+\d\theta)$ versus $\theta$ in Figure \ref{fig:t}, with reference to the value of $\t(\z,\Y)=3.85$. Figure~\ref{fig:t} shows that the basic equation $\t(\z,\Y)=\t(\zp,\Y+\d\theta)$ has two roots $\theta_{\pi}=(1.33,2.27)$.
\end{example}

\subsection{Step 2: Determining changes at jump points}
\label{sec:2}

When the $p$-value function is monotonic, the directions of changes of the $p$-value function at all jump points are the same, which is not true for the non-monotonic $p$-value function produced by the studentized $t$-statistic. 
Even though we need to determine the direction of change for each jump point individually, this is not a difficult task.
For every single $\zp$, there are only three types of changes for the relationship between $\t(\zp,\tY)$ and $\t(\z,\Y)$: (i) from $\t(\zp,\tY)<\t(\z,\Y)$ to $\t(\zp,\tY)>\t(\z,\Y)$, represented by $J_\pi=1$; (ii) from $\t(\zp,\tY)>\t(\z,\Y)$ to $\t(\zp,\tY)<\t(\z,\Y)$, represented by $J_\pi=-1$; (iii) the relationship remains unchanged, represented by $J_\pi=0$. Thus, to determine $J_\pi$, we can evaluate $\t(\zp,\tY)$ for $\theta=\theta_\pi\pm \epsilon$, where $\epsilon>0$ needs to be less than the spacing between all roots.

Since there might be cases where $\zp$ and $\zpp$ have equal roots $\theta_\pi=\theta_{\pi'}$, we aggregate these $J_\pi$'s corresponding to $\theta_\pi$, denoted by $(\theta_\pi, J_\pi)$.
Consequently, $J_\pi$ may be equal to numbers such as $\pm 2$, $\pm3$, and so on.
We remove those $\theta_\pi$ with $J_\pi=0$, which are not actually jump points.
Finally, we denote the number of jump points by $K$ and sort $(\theta_\pi, J_\pi)$ by $\theta_\pi$ from small to large, denoted by $(\theta_{(k)}, J_{(k)})$, $k=1,\ldots,K$.

\begin{examplecont}
For $\zp=(1,1,1,1,0,0,0,0)$, we can determine that $J_{\pi,1}=1$ for the root $\theta_{\pi,1}=1.33$ and $J_{\pi,2}=-1$ for the root $\theta_{\pi,2}=2.27$, as shown in Figure \ref{fig:t}. 
\end{examplecont}

\subsection{Step 3: Recovering \textit{p}-value function}

Now, we can recover the the $p$-value function using $(\theta_{(k)}, J_{(k)})$. We first consider the $p$-value function for $H_1^{\theta+}:Y_i(1)-Y_i(0)>\theta$.

We calculate the $p$-value before the first jump point by $p_0=p^+(\theta_{(1)}-\epsilon)$, where $\epsilon>0$ is arbitrary. Then, we iteratively determine the $p$-values at each jump point $\theta_{(k)}$ by
$$
p_k = p_{k-1} + J_{(k)} / |\cZ|, \; k=1,\ldots,K,
$$
where $1/|\cZ|$ is the jump height that every single $\zp$ contributes to the $p$-value function. 
Then, the $p$-value function is recovered by all jump points $\theta_{(k)}$ and $p_k$.
To obtain a whole $p$-value function conveniently, we can use \texttt{R} function \texttt{stepfun} with $\theta_{(k)}$ and $p_k$.

For $H_1^{\theta-}:Y_i(1)-Y_i(0)<\theta$, we need to use $p_0=p^-(\theta_{(1)}-\epsilon)$ and iterative formula
$$
p_k = p_{k-1} - J_{(k)} / |\cZ|,\; k=1,\ldots,K.
$$

\begin{figure}[p]
\centering
    \includegraphics[width = 0.75\textwidth]{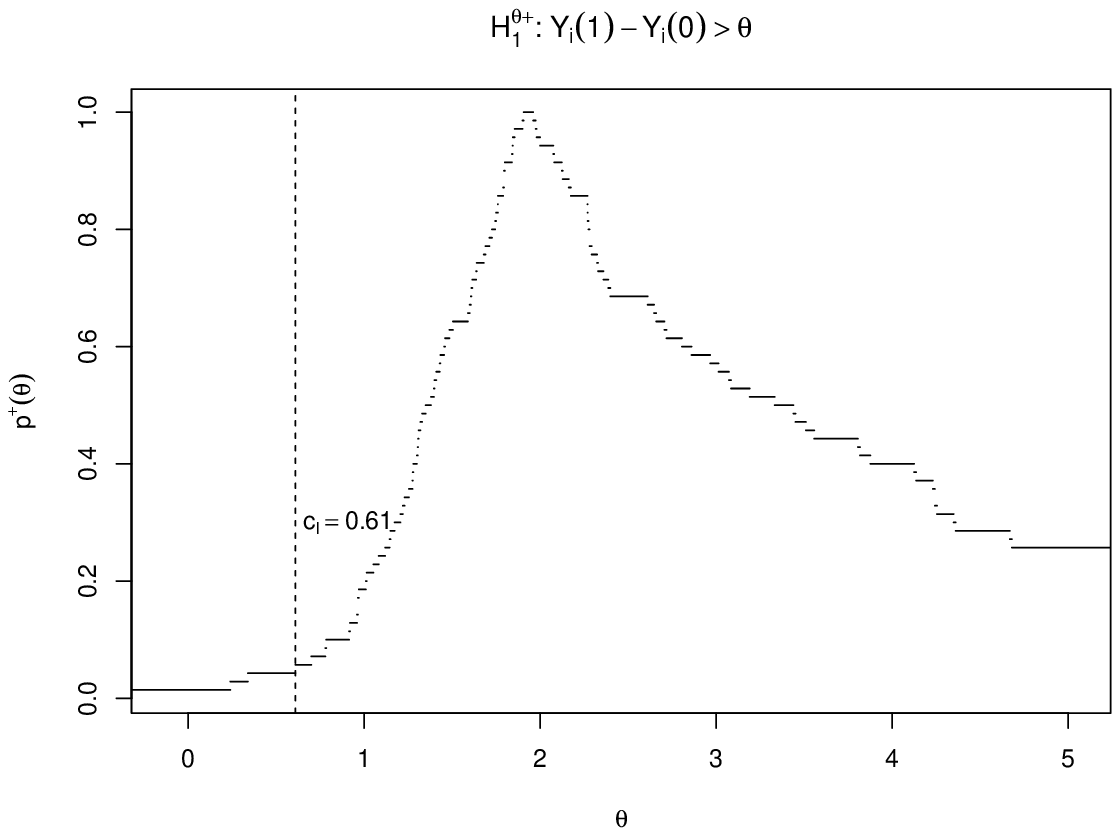}
\caption{Non-monotonic $p$-value function recovered by the proposed analytic approach. The vertical dashed line is the 95\% lower confidence bound. This Figure corresponds to Example \ref{ex:1} with 8 units.}
\label{fig:p}
\end{figure}

\begin{examplecont}
Consider $H_1^{\theta+}:Y_i(1)-Y_i(0)>\theta$. We use the proposed analytic approach to recover the non-monotonic $p$-value function, as shown in Figure \ref{fig:p}.
\end{examplecont}

\subsection{Step 4: Squeezing confidence intervals from \textit{p}-value function}
\label{sec:4}

\citet{luo2021} uses the bisection method to solve $p^+(\theta)=\alpha$, given that the $p$-value function is monotonic. When the $p$-value function is non-monotonic, their method cannot guarantee the desired coverage probability. In contrast, exploiting the $p$-value function recovered in Step 3, we can squeeze confidence intervals with the desired coverage probability.

We first consider the $1-\alpha$ lower confidence bound for $H_1^{\theta+}:Y_i(1)-Y_i(0)>\theta$. We start with $c_l=-\infty$ and compare $p_0$ with $\alpha$. If $p_0 < \alpha$, let $c_l=\theta_{(1)}$. Then, if $p_1 < \alpha$, let $c_l=\theta_{(2)}$. We squeeze the confidence bound sequentially until $c_l=\theta_{(k)}$ with $p_k> \alpha$ for some $k\leq K$. Our procedure naturally ensure that $p^+(\theta)< \alpha$ for all $\theta<c_l$, thus, $[c_l,\infty)$ is theoretically guaranteed to achieve the desired coverage probability. Since the $p$-value function is non-monotonic, we do not squeeze out all set of $\theta$ such that $p^+(\theta)\le \alpha$. It is also feasible to do so, which will generate a confidence set with discontinuous intervals, potentially making it less favorable in practice.

For $H_1^{\theta-}:Y_i(1)-Y_i(0)<\theta$, we can obtain the $1-\alpha$ upper confidence bound similarly. For $H_1^{\theta\pm}:Y_i(1)-Y_i(0)\neq\theta$, we can obtain the $1-\alpha$ confidence interval by solving the $1-\alpha/2$ lower and upper confidence bounds, respectively.

\begin{examplecont}
Consider $H_1^{\theta+}:Y_i(1)-Y_i(0)>\theta$. Using the proposed analytic approach, we can obtain the 95\% lower confidence bound $c_l=0.61$, as shown in Figure \ref{fig:p}.
\end{examplecont}

\section{Simulation}
\label{sec:sim}

We conduct simulations to confirm our findings.
We first consider the data set in Example \ref{ex:1} with sample size $n=8$. Then, we consider another case with $n=100$ and potential outcomes being generated by $Y_i(0)\sim N(0,1)$ and $Y_i(1)=Y_i(0)+1$. We randomly assign $n/2$ units to the treatment group and $n/2$ units to the control group. We use studentized $t$-statistic as the test statistic to perform FRTs and construct RBCIs. We consider two-sided alternative hypothesis with $\alpha=0.05$.

In Example \ref{ex:1}, there are 70 assignments in $\cZ$. Thus, we can enumerate all assignments to perform randomization-based inference. Using the proposed method, generating one RBCI takes about 0.3 seconds. For reference, generating one exact $p$-value takes about 0.006 seconds. The coverage probability is 95.7\% and the type I error is 0.0286. 
The coverage probability of RBCI exceeds the nominal level due to the discreteness of the $p$-value function, especially when the sample size is very small. In Example \ref{ex:1} with only 8 units, the $p$-value function $p^+(\theta)$ jumps from 0.043 to 0.057 at $c_l=0.61$. Thus, the exact coverage probability is $1-0.043=0.957$, exceeding the nominal level of $0.95$. In general, the exact coverage probability of an RBCI, $[c_l,c_u]$, is $1-\max_{\theta\notin [c_l,c_u]}p(\theta)$, which is equal to or greater than the nominal level due to our construction method.

When $n=100$, there are about $10^{29}$ assignments in $\cZ$. Thus, we use Monte Carlo to generate $n_{\text{fisher}} = 10^4$ assignments for randomization-based inference and $n_{\rm rep}=10^3$ replications to evaluate repeated sampling performance. Using the proposed method, generating one two-sided RBCI takes about 23 seconds. For reference, generating one exact $p$-value takes about 0.8 seconds. The coverage probability is 95.4\% and the type I error is 0.046.
Since the $p$-value function is not guaranteed to be monotonic for studentized $t$-statistic, the bisection method is not applicable \citep{luo2021}. 
We consider the grid method for comparison. Based on the discussion in Section \ref{sec:1}, if every $\zp$ has two roots, it is expected that $p^+(\theta)$ will have $2\times10^4$ jump points. For one-sided RBCI, it takes about 4 hours to evaluate $p^+(\theta)$ at $2\times10^4$ even-spaced grids. However, since jump points are not actually even-spaced, the RBCI based on these $2\times10^4$ grids is still not guaranteed to have the desired coverage probability. Thus, the grid method is infeasible.

In the simulation conducted by \citet{fiksel2024}, the additional time required to construct the RBCI is negligible when compared to computing the Fisher exact $p$-value. This is because when the $p$-value function is monotonic, ordering the jump points suffices to recover the $p$-value function and derive the RBCI. However, in scenarios where the $p$-value function is non-monotonic, we need to squeeze through each jump point to obtain the RBCI, which takes longer.  Therefore, it is advisable to determine the monotonicity of the $p$-value function beforehand. If it is monotonic, the ordering approach is recommended \citep{zhu2023pair,fiksel2024}. Conversely, if the $p$-value function is non-monotonic, employing the squeezing approach becomes necessary.

We also examine the statistical and computational performance of the proposed method for different sample sizes $n=50,100,500,1000,5000,10000$ with different numbers of Monte Carlo $n_{\text{fisher}}=10^4, 2\times 10^4$. We simulate for $n_{\rm rep}=10^3$ replications to evaluate repeated sampling performance. Table \ref{tab:sim} shows the simulation results. 
The RBCI achieves the nominal level, and the type I error is under control across all sample sizes. With increasing sample sizes, the computational time increases, but remains acceptable. When the number of Monte Carlo doubles, the computational time also doubles. However, the statistical performance is similar, indicating that $n_{\text{fisher}} = 10000$ is sufficient for our simulation setup.
Parallel computing with fast implementation such as \texttt{Rcpp} could further reduce the computational time.

\section{Discussion}

In this rejoinder, we propose an analytical approach to invert the FRT to construct confidence intervals for non-monotonic test statistics. Steps 2, 3, and 4 are applicable to all non-monotonic statistics. 
In contrast, solving jump points in step 1 requires a case-by-case analysis for various test statistics.
Given that \citet{fiksel2024} has provided a jump point solution for the covariate-adjusted test statistic, applying steps 2 to 4 enables us to generate randomization-based covariate-adjusted confidence intervals with the desired coverage probability, eliminating the need for the monotonic condition.

Solving jump points for other test statistics opens a new direction for future research. For complex test statistics, if the closed-form solution for jump points are not available, numerical methods can be involved. 
Nevertheless, in terms of both theoretical guarantee and computational efficiency, it is beneficial to solve $T(\z,\Y)=T(\zp,\boldsymbol{Y}+\d\theta)$ rather than solving $p^+(\theta)=\alpha$. The proposed approach is applicable to any experimental design, although we still need to solve jump points for test statistics tailored for different designs. 

FRTs have deep connections with many other statistical methods, such as permutation tests, conformal prediction, and confidence distribution, as revealed by \citet{zhang2023randomization} and \citet{luo2021}. Applying the proposed analytic inversion approach within these areas is also promising.

\section*{Acknowledgements}

Hanzhong Liu's research is partially supported by the National Natural Science Foundation of China (No. 12071242). 
Ke Zhu acknowledges the financial support from Dr. Shu Yang at North Carolina State University and Dr. Xiaofei Wang at Duke University during his postdoctoral research. 

\section*{Supplementary Materials}
R codes for implementing the methods and reproducing the numerical results are available with this paper at the Biometrics website on Oxford Academic.

\section*{Data Availability}
Data sharing is not applicable.

\bibliographystyle{apalike}
\bibliography{causal.bib}

\newpage

\begin{table}[p]

\caption{\label{tab:sim}Simulation results for different $n$ and $n_{\text{fisher}}$.}
\centering
\begin{threeparttable}
\begin{tabular}[t]{rccccc}
\toprule
$n$ & $n_{\text{fisher}}$ & Coverage & Type I Error & Time for $p$-value (sec.) & Time for RBCI (sec.)\\
\midrule
50 & 10000 & 0.963 & 0.037 & 0.7 & 22.5\\
50 & 20000 & 0.951 & 0.049 & 1.4 & 45.0\\
\midrule
100 & 10000 & 0.954 & 0.046 & 0.8 & 23.2\\
100 & 20000 & 0.958 & 0.042 & 1.5 & 46.2\\
\midrule
500 & 10000 & 0.944 & 0.056 & 1.0 & 25.6\\
500 & 20000 & 0.940 & 0.060 & 2.1 & 51.8\\
\midrule
1000 & 10000 & 0.944 & 0.056 & 1.3 & 28.3\\
1000 & 20000 & 0.956 & 0.044 & 2.8 & 56.5\\
\midrule
5000 & 10000 & 0.944 & 0.056 & 5.1 & 62.1\\
5000 & 20000 & 0.959 & 0.041 & 9.5 & 122.4\\
\midrule
10000 & 10000 & 0.949 & 0.051 & 13.0 & 146.4\\
10000 & 20000 & 0.965 & 0.035 & 28.0 & 282.4\\
\bottomrule
\end{tabular}
\begin{tablenotes}
\item Note: Coverage is the empirical coverage probability in 1000 replications; Time for $p$-value (sec.) (Time for RBCI (sec.)) is the time required for calculating one $p$-value (RBCI) on a personal laptop, not the time required for 1000 replications.
\end{tablenotes}
\end{threeparttable}
\end{table}

\end{document}